\documentclass[aps,prl,epsfig,showpacs,floats,twocolumn,superscriptaddress,amssymb,amsmath,floatfix]{revtex4}
\usepackage{graphicx}
\begin{document}

\title{Generic Conditions for Hydrodynamic Synchronization}

\author{Nariya Uchida}
\email{uchida@cmpt.phys.tohoku.ac.jp}
\affiliation{Department of Physics, Tohoku University, Sendai, 980-8578, Japan}

\author{Ramin Golestanian}
\email{ramin.golestanian@physics.ox.ac.uk}
\affiliation{The Rudolf Peierls Centre for Theoretical Physics, University of Oxford,\\
1 Keble Road, Oxford OX1 3NP, United Kingdom}

\date{\today}

\begin{abstract}
Synchronization of actively oscillating organelles such as
cilia and flagella facilitates self-propulsion of cells
and pumping fluid in low Reynolds number environments.
To understand the key mechanism behind synchronization
induced by hydrodynamic interaction, we study a model of
rigid-body rotors making fixed trajectories of arbitrary shape
under driving forces that are arbitrary functions of the phases.
For a wide class of geometries, we obtain the necessary and
sufficient conditions for synchronization of a pair of rotors.
We also find a novel synchronized pattern with a time-dependent 
phase shift.
Our results shed light on the role of
hydrodynamic interactions in biological systems, and could
help in developing efficient mixing and transport strategies
in microfluidic devices.
\end{abstract}


\def\commentoff#1{}
\def\commenton#1{{\sf #1}}

\maketitle

\paragraph{Introduction.}

The idea that hydrodynamic interactions at low Reynolds number can induce
synchronization between active components with cyclic motion has
been the subject of extensive studies
since the pioneering work of G. I. Taylor~\cite{Taylor},
and has culminated in a number of direct experimental demonstrations in
recent years~\cite{exp,Goldstein,QJ09}. For example, the effective and recovery strokes
of beating cilia~\cite{Blake-Sleigh} are considered to be important for generating
their coordinated motion (metachrony)~\cite{GL97,KN06,GJ07}. While resolving
the intricate conformations of the elastic filaments is important for studying
coordination in high density assemblies, it can be argued that at sufficiently
low densities hydrodynamic interaction does not alter the beating pattern of
the active filaments, such that they can be feasibly modeled as simple
beads following fixed trajectories~\cite{VJ06,RL06}.
The simplicity of this level of description allows for
complex many-body effects to be probed in large arrays
of such beads with additional active internal mechanisms~\cite{GJ07,LJB03,UG10}.

There have been a number of recent studies on hydrodynamic interaction between 
rotating or orbiting rigid bodies~\cite{VJ06,RL06,KP04,NEL08,RS05}.
An emerging general picture suggests that 
rigid bodies making fixed trajectories do not easily synchronize.
Rigid helices with parallel axes~\cite{KP04} or beads on circular
trajectories~\cite{RL06} with constant driving torque do not synchronize,
unless flexibility is introduced in the orientation of the rotation axis~\cite{RS05}
or in the confinement to the trajectory~\cite{NEL08}, respectively.
Vilfan and J\"ulicher~\cite{VJ06} studied two beads on tilted elliptic
trajectories near a substrate, with a velocity-dependent driving force.
They found that both the height-dependence of the drag coefficient
and the eccentricity of the trajectories are necessary to stabilize the synchronized
state.
Ryskin and Lenz~\cite{RL06} considered a more general
model, in which each cilium is represented by a collection of beads
connected to each other.
Each bead makes a fixed trajectory of arbitrary shape under
a driving force that is an arbitrary function of the phase.
They applied the general framework to
a variety of beating patterns that mimic the ciliary strokes,
and found them to be able to stabilize traveling (metachronal) waves
but not synchronized states.
These results naturally lead to the following question:
when do objects with fixed trajectories synchronize via hydrodynamic
interaction?
Here, we address this question by formulating generic and explicit
criteria for hydrodynamic synchronization.

We use a simple version of Ryskin-Lenz model
in which each active object (rotor) is made of a single bead.
We derive a necessary and sufficient condition
for a pair of rotors to synchronize,
in terms of the trajectory shape and force profile.
We apply the obtained criterion to specific 
trajectories, and identify the form of the force profiles
that cause synchronization.
For circular trajectories, for example,
we find the requirement that
the {\it logarithm} of the force has a non-vanishing second-harmonic
component of a specific sign, which originates from
the second-rank tensorial nature of hydrodynamic interaction.
We consider trajectories in the bulk and near a substrate,
and those tilted relative to each other.
We also develop an effective potential picture 
to examine the global stability of the synchronized states,
which reveals a novel synchronized pattern with a time-dependent phase shift.

\paragraph{Dynamical equations.}

\begin{figure}[t]
\includegraphics[width=0.99\columnwidth]{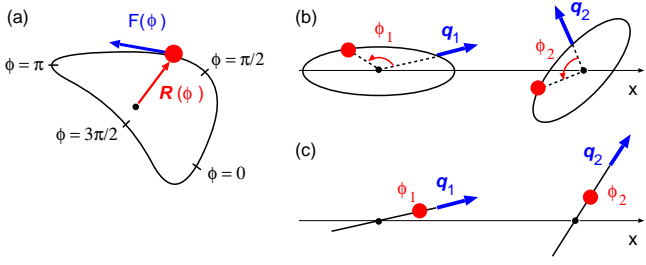}
\caption{
(a) A generic trajectory with its shape specified by ${\bf R}(\phi)$.
The bead is driven by the tangential force $F(\phi)$.
(b) Circular trajectories ${\bf Q}_1 \cdot {\bf R}(\phi)$ and ${\bf Q}_2 \cdot {\bf R}(\phi)$
with ${\bf R}(\phi) = b (\cos\phi, \sin\phi, 0)$ and the
rotation matrices ${\bf Q}_1, {\bf Q}_2$. Their orientations are specified by
the unit vectors ${\bf q}_i = {\bf Q}_i \cdot {\bf e}_x$.
(c) Linear trajectories with ${\bf R}(\phi) = R(\phi) {\bf e}_x$ and
their orientations specified by ${\bf q}_i = {\bf Q}_i \cdot {\bf e}_x$.
The centers of the trajectories are both on the $x$-axis.
\label{fig:twobeads}
}
\end{figure}

We consider a pair of rotors (indexed by $i=1,2$) 
and assume that each is a spherical bead of radius $a$ 
that follows a fixed periodic trajectory
${\bf r}_i = {\bf r}_i(\phi_i)$, where $\phi_i = \phi_i(t)$ is the phase variable
with the period $2\pi$ [see Fig. \ref{fig:twobeads}(a)].
The bead is driven by an active force $F_i = F_i(\phi_i)$
that is tangential to the orbit and is an arbitrary function of the phase.
The hydrodynamic drag force acting on the bead is given by
${\bf g}_i = \zeta [{\bf v}({\bf r}_i) - \dot{\bf r}_i]$,
where $\zeta = 6\pi \eta a$ is the drag coefficient~\cite{note_on_zeta},
and ${\bf v}({\bf r})$ is the velocity field of the surrounding fluid.
The tangential component of the drag force
is balanced by the driving force acting on each rotor, namely,
\begin{math}
F_i + {\bf t}_i \cdot {\bf g}_i = 0,
\end{math}
where ${\bf t}_i$ is the tangential unit vector of the orbit given by
${\bf t}_i = {{\bf r}'_i}/{|{\bf r}'_i|}$ with ${\bf r}'_i = {d{\bf r}_i}/{d\phi_i}$.
Substituting the expression for the drag force with
$\dot{\bf r}_i =  {\bf r}'_i \dot\phi_i$ into the force balance equation, we
obtain the phase velocity as
\begin{math}
\dot\phi_i
= \omega_i
+ {{\bf t}_i \cdot {\bf v}({\bf r}_i)}/{|{\bf r}'_i|},
\end{math}
where
\begin{math}
\omega_i(\phi_i)
= {F_i(\phi_i)}/{\zeta |{\bf r}_i'|}
\end{math}
is the intrinsic phase velocity.
The reaction force $- {\bf g}_i$ exerted by the bead on the fluid
generates the flow field
\begin{eqnarray}
{\bf v}({\bf r}) = - \sum_j {\bf G}({\bf r}, {\bf r}_j) \cdot {\bf g}_j
\simeq
\sum_{j} \zeta {\bf G}({\bf r}, {\bf r}_j) \cdot {\bf r}'_j \omega_j,
\label{vflow}
\end{eqnarray}
where ${\bf G}({\bf r}, {\bf r}_j)$ is the Oseen tensor
describing the hydrodynamic interaction in bulk fluid.
On the RHS of Eq. (\ref{vflow}), we assumed $|{\bf r} - {\bf r}_j| \gg a$
and retained the leading order term
with respect to
$\zeta {\bf G}({\bf r}, {\bf r}_j) = O(a/|{\bf r}-{\bf r}_j|)$.
Substituting this into the above expression for the phase velocity,
we arrive at the coupled phase oscillator equation
\begin{eqnarray}
\dot\phi_i
= \omega_i +
\sum_{j\neq i}
\left(
\frac{{\bf t}_i}{|{\bf r}'_i|}
\cdot \zeta {\bf G}_{ij} \cdot |{\bf r}'_j| {\bf t}_j
\right)
\omega_j,
\label{Dotphi1}
\end{eqnarray}
where ${\bf G}_{ij} = {\bf G}({\bf r}_i, {\bf r}_j)$.

We now assume that the two trajectories have the same shape but are
oriented differently relative to the axis that connects their centers.
We can write each trajectory as ${\bf r}_i(\phi) = {\bf r}_{i0} + {\bf Q}_i \cdot {\bf R}(\phi)$,
where ${\bf r}_{i0}$ is the position of the center, ${\bf R}(\phi)$ describes the shape
of the trajectory, and ${\bf Q}_i$ is a rotation matrix.
We also assume that the center positions are
on the $x$-axis and are separated by the distance $d$ ($\gg a$)
from each other,
${\bf r}_{10} = (0,0,0)$ and ${\bf r}_{20} = (d,0,0)$.
Using ${\bf r}'_i(\phi) = |{\bf R}'(\phi)| {\bf Q}_i \cdot{\bf t}(\phi)$
with the unit vector ${\bf t}(\phi) = {\bf R}'(\phi)/|{\bf R}'(\phi)|$
in Eq. (\ref{Dotphi1}),
we obtain the difference between the phase velocities as
\begin{eqnarray}
\dot{\phi}_1 -  \dot{\phi}_2 &=& \omega(\phi_1) - \omega(\phi_2)
\nonumber\\
&+&
\left[
\frac{F(\phi_2)}{F(\phi_1)} \omega(\phi_1)
-
\frac{F(\phi_1)}{F(\phi_2)} \omega(\phi_2)
\right]
H(\phi_1, \phi_2),
\quad
\label{dotdelta}
\end{eqnarray}
where
\begin{eqnarray}
\omega(\phi_i) 
= \frac{F(\phi_i)}{\zeta |{\bf R}'(\phi_i)|}.
\end{eqnarray}
is the intrinsic phase velocity, and
we have introduced the coupling function
\begin{eqnarray}
H(\phi_1, \phi_2)
&=&
{\bf Q}_1 \cdot {\bf t}(\phi_1)
\cdot \zeta {\bf G}_{12} \cdot
{\bf Q}_2 \cdot {\bf t}(\phi_2),
\label{Hdef}
\end{eqnarray}
which is a dimensionless quantity of order $O(a/d)$.
To examine the stability of the synchronized state,
we set $\phi_1 = \phi(t) + \delta(t)$, $\phi_2 = \phi(t)$
and linearize Eq. (\ref{dotdelta}) with respect to
the phase difference $\delta(t)$, which gives
the linear growth rate
\begin{eqnarray}
\frac{\dot\delta}{\delta}
=
\omega'(\phi) +
\left[
\omega'(\phi) - \frac{2 F'(\phi)}{F(\phi)} \omega(\phi)
\right]
H(\phi, \phi).
\end{eqnarray}
Integrating the above result over the period $T = \int_0^{2\pi} d\phi/\dot\phi$,
we obtain the cycle-averaged growth rate as
\begin{eqnarray}
\Gamma
= -\frac{2}{T} \int_0^{2\pi} d\phi \, [\ln F(\phi)]' H(\phi, \phi),
\label{stabcond}
\end{eqnarray}
to the lowest order in the hydrodynamic coupling $H$.
A stable synchronized state exists when $\Gamma<0$. Equation (\ref{stabcond}) thus shows
that a necessary condition for synchronization is that {\em both} the force profile $F(\phi)$
and the hydrodynamic coupling $H(\phi,\phi)$ are not constant.
For any given trajectory ${\bf R}(\phi)$
that gives a non-constant function $H(\phi, \phi)$,
we can prescribe a force profile $F(\phi)$
that satisfies the above condition.
For example, the force profile
$
F(\phi) =
F_0 \left[
1 + \int_0^\phi d\psi \left(H(\psi,\psi) - \overline{H}\right)
\right]$, with $\overline{H}$ being the cycle average of $H(\phi,\phi)$,
makes $\Gamma$ negative-definite to
the leading order in the coupling.

In order to calculate $H(\phi,\phi)$,
we decompose the Oseen tensor into
isotropic (I) and dyadic (D) parts as
\begin{eqnarray}
\zeta {\bf G}_{12}
&=&
G_I(r_{12}) {\bf I} + G_D(r_{12}) \frac{{\bf r}_{12} {\bf r}_{12}}{r_{12}^2},
\label{G12}
\end{eqnarray}
where $G_I(r) = G_D(r) = 3/4r$ and we have used
\begin{eqnarray}
{\bf r}_{12}
= {\bf r}_1 - {\bf r}_2
= -d {\bf e}_x
+ {\bf Q}_1 \cdot {\bf R}(\phi_1) - {\bf Q}_2 \cdot {\bf R}(\phi_2).
\qquad
\label{r12}
\end{eqnarray}
When the characteristic dimension $b=\max |{\bf R}(\phi)|$
of the trajectory is much smaller than the distance,
we can approximate the hydrodynamic interaction kernel as
$
\zeta G_{12} \simeq G_I(d) {\bf I} + G_D(d) {\bf e}_x {\bf e}_x.
\label{G12approx}
$
Under this approximation, the coupling function (\ref{Hdef})
becomes
\begin{eqnarray}
H(\phi,\phi) &=& G_D(d)\,[{\bf q}_1 \cdot {\bf t}(\phi)] \,[{\bf q}_2 \cdot {\bf t}(\phi)]
+ {\rm const},
\label{H0th}
\end{eqnarray}
where ${\bf q}_i = {\bf Q}_i \cdot {\bf e}_x$.
Note that the diagonal part of the hydrodynamic kernel
gives a constant contribution to $H(\phi,\phi)$
and hence drops off from the integral (\ref{stabcond}).
We now examine a number of cases in more details.

\paragraph{Circular trajectories.}

\begin{figure}[t]
\includegraphics[width=0.74\columnwidth]{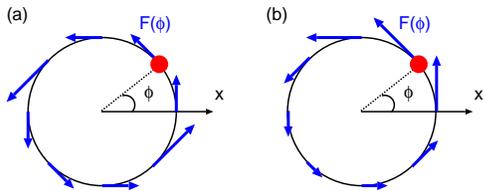}
\caption{
Examples of the force profiles that act to synchronize
two beads on circular trajectories aligned on the $x$-axis.
(a) $F(\phi)=  F_0 [1 - \frac12 \sin (2\phi)]$. 
(b) $F(\phi)=  F_0 \left[1 + \frac12 
\sin \left(\phi + \frac{\pi}{4}\right)\right]$.
\label{fig:circles}
}
\end{figure}

As the first example, let us consider the circular trajectory [see Fig. \ref{fig:twobeads}(b)]
\begin{eqnarray}
{\bf R}(\phi) = b (\cos\phi, \sin\phi, 0).
\label{circle}
\end{eqnarray}
For this trajectory,  we have
$|{\bf R}'(\phi)| = b$
and ${\bf t}(\phi) = (-\sin\phi, \cos\phi, 0)$.
First we consider mutually parallel trajectories with ${\bf Q}_1 = {\bf Q}_2 = {\bf I}$.
In this case, we have ${\bf q}_i = {\bf e}_x$ and
$H(\phi,\phi)
= G_D(d) \sin^2\phi
= -\frac{1}{2} G_D(d) \cos(2\phi) + {\rm const}$.
Note that the factor $\cos 2\phi$ represents the second-rank
tensorial nature of the hydrodynamic kernel.
To stabilize the synchronized state,
we can use the force profile
$F(\phi)=  F_0 [1 - A \sin (2\phi)] \, (F_0 > 0, 0 < A < 1)$,
which gives $\Gamma = -\frac{G_D(d)}{T} A + O(A^2) < 0$
[see Fig. \ref{fig:circles}(a) for illustration].
We can also use
$F(\phi)=  F_0 \left[1 + B\sin \left(\phi + \frac{\pi}{4}\right)\right] \,
(F_0 > 0, -1 < B <1)$,
which gives $\Gamma = - \frac{G_D(d)}{2T} B^2 + O(B^4)< 0$
[see Fig. \ref{fig:circles}(b)].
In general, the synchronized state
is linearly stable if and only if
the Fourier expansion of $\ln F (\phi)$
has a negative coefficient for $\sin 2\phi$.

Next, we consider rotated circular trajectories.
For each trajectory ($i=1,2$), the rotation operator that acts on (\ref{circle})
is parameterized by the Euler angles $(\alpha_i, \beta_i, \gamma_i)$ as
$
{\bf Q}_i =
{\bf M}_z(\gamma_{i})
{\bf M}_x(\beta_{i})
{\bf M}_z(\alpha_{i}),
\label{Euler}
$
where ${\bf M}_x(\theta)$ and ${\bf M}_z(\theta)$
are the matrices of rotation by angle $\theta$ around the $x$ and $z$-axis, respectively.
It gives
$
{\bf q}_i =
(\cos\alpha_i \cos\gamma_i - \cos\alpha_i \cos\beta_i \sin\gamma_i,
-\sin\alpha_i \cos\gamma_i - \cos\alpha_i \cos\beta_i \cos\gamma_i,
\sin\beta_i \sin\gamma_i).
$
For a rotation in the $xy$-plane ($\beta_i = \gamma_i=0$),
we get
$H(\phi,\phi) = - \frac{1}{2} G_D(d)
\cos(2\phi - \alpha_1 - \alpha_2) + {\rm const}.$,
and synchronization is induced by, for example, the force profile
$
F(\phi)= F_0 [1 - A \sin (2\phi -\alpha_1 - \alpha_2)] \, (0 < A < 1)
$.
Next, a rotation in the $yz$-plane ($\alpha_i = \gamma_i = 0$)
gives
$H(\phi,\phi) = - \frac{1}{2} G_D(d)
\cos \beta_1 \cos \beta_2 \cos(2\phi) + {\rm const}$.
Finally, circular trajectories
that are vertical to the $xy$-plane
($\alpha_i = 0, \beta_i = \frac{\pi}{2}$),
give
$
H(\phi,\phi) = -\frac{1}{2} G_D(d) \cos \gamma_1 \cos \gamma_2 \cos(2\phi)
+ {\rm const}.
$
All of these cases yield similar conditions for synchronization
in terms of the second Fourier coefficients of $\ln F (\phi)$,
as in the case of non-rotated circular trajectories discussed above. Note that
the decay rate to a synchronized state is independent of the size of the trajectory
$b$ at the leading order for circular trajectories.

\paragraph{Linear trajectories.}

Next we consider the linear trajectory [see Fig. \ref{fig:twobeads}(c)]
\begin{eqnarray}
{\bf R}(\phi) = R(\phi) {\bf e}_x,
\end{eqnarray}
that gives ${\bf t}(\phi) = {\rm sgn} [R'(\phi)] {\bf e}_x$, which amounts
to a constant contribution to the coupling function (\ref{H0th})
and hence neither stabilize nor destabilize the synchronized state.
However, $O(b/d)$ corrections to the hydrodynamic kernel
give non-constant contributions.
Substituting Eq. (\ref{r12}) into (\ref{G12}) and (\ref{Hdef}),
and retaining the first order term
with respect to $R(\phi)$,
\commentoff{
we have
\begin{eqnarray}
\zeta {\bf G}_{12}
&=&
\left[G_I(d) + G_I'(d) p_x R(\phi) \right] {\bf I}
+
\left[G_D(d) + G_D'(d) p_x R(\phi) \right] {\bf e}_x {\bf e}_x
+
\frac{G_D(d)}{d} R(\phi)
\left(
{\bf e}_x {\bf p} + {\bf p} {\bf e}_x
\right),
\nonumber\\
\end{eqnarray}
where ${\bf p} = {\bf q}_1 - {\bf q}_2 = ({\bf Q}_1 - {\bf Q}_2) \cdot {\bf e}_x$.
It gives
}
we obtain
the coupling function
\begin{eqnarray}
H(\phi,\phi) &=& -2 R(\phi)
\Bigg[
G_I'(d) ({\bf q}_1 \cdot {\bf q}_2) p_x
+
G_D'(d) q_{1x} q_{2x} p_x
\nonumber\\
&+& \frac{G_D(d)}{d}
\left(
q_{1x} {\bf q}_2 \cdot {\bf p} + q_{2x} {\bf q}_1 \cdot {\bf p}
\right)
\Bigg]
+ {\rm const},
\end{eqnarray}
where ${\bf p} = {\bf q}_1 - {\bf q}_2$.
Note that the coupling is constant when ${\bf q}_1 = {\bf q}_2$,
because the distance between the two beads is constant
when the two trajectories are parallel.
When they are not parallel, the phase dependence
is proportional to $R(\phi)$.
For example, for
perpendicular trajectories with ${\bf q}_1 = {\bf e}_x$, ${\bf q}_2 = {\bf e}_y$
and the orbital profile $R(\phi) = b\cos\phi$,
the synchronized state is stabilized if and only if
the Fourier expansion of $\ln F(\phi)$ has
a positive coefficient for $\sin \phi$.

\paragraph{Nonlinear stability analysis.}

\begin{figure}[t]
\includegraphics[width=0.99\columnwidth]{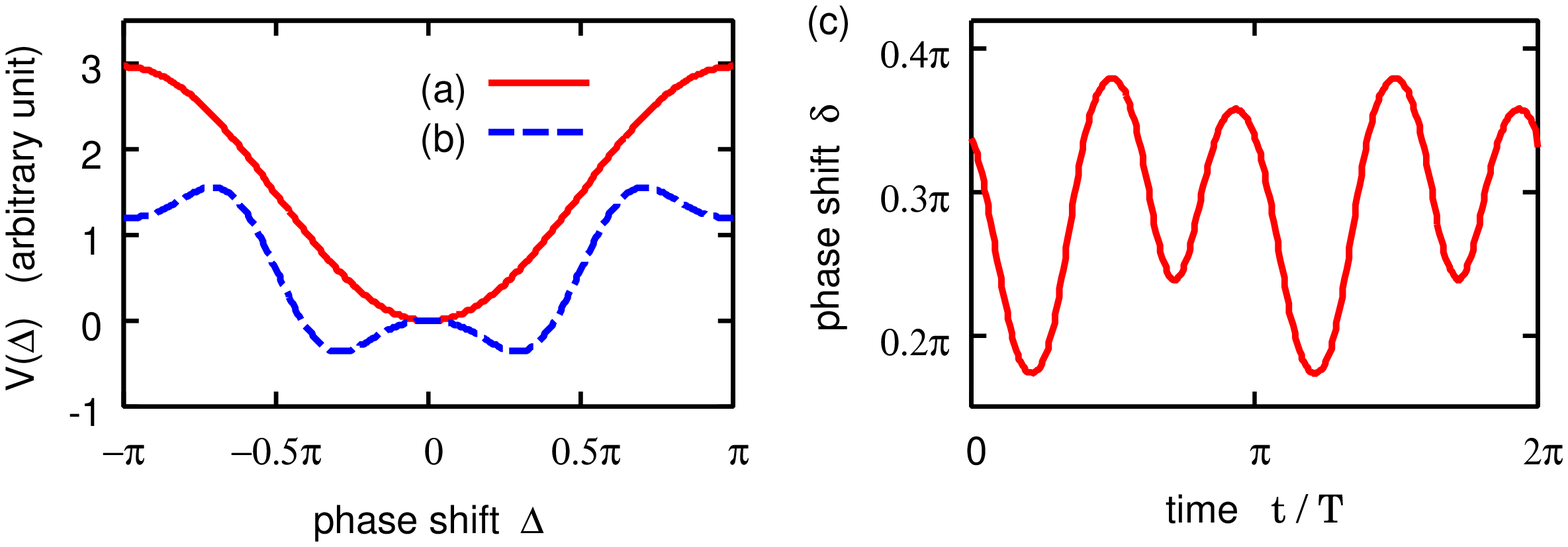}
\caption{
Examples of the effective potential $V(\Delta)$.
(a) For the circular trajectory with $F(\phi)=  F_0 [1 - \frac12 \sin (2\phi)]$,
the unique stable solution is found at $\Delta=0$.
(b) For the elliptic trajectory ${\bf R}(\phi) = b(\cos\phi, \frac12 \sin\phi,0)$
with $F(\phi)=  F_0 [1 - \frac{1}{10} \sin (2\phi) + \frac12 \sin(4\phi)]$,
we obtain the bistable solutions $\Delta = \pm \Delta_0$ with
$\Delta_0 \simeq 0.29 \pi$, and a metastable solution at $\Delta=\pi$.
(c) The stable solution $\Delta = \Delta_0$ in (b) gives a time-dependent
phase shift $\delta = \delta(t)$ in the original gauge.
\label{fig:VofDelta}
}
\end{figure}

We can analyze global stability of the synchronized state
by a nonlinear evolution equation for the phase difference.
To derive it, first we reparameterize the trajectory by
the new phase variable $\Phi = \Phi(\phi)$ that makes
the intrinsic phase velocity constant:
$\Phi'(\phi) \cdot F(\phi)/\zeta |{\bf R}'(\phi)|= 2\pi/T$.
In this gauge, the phase difference
$\Delta = \Phi_1 - \Phi_2 = \Phi(\phi_1) - \Phi(\phi_2)$
obeys [see Eq. (\ref{dotdelta})]
\begin{eqnarray}
\dot\Delta = \frac{2\pi}{T}
\left[
\frac{\tilde{F}(\Phi_2)}{\tilde{F}(\Phi_2 + \Delta)} -
\frac{\tilde{F}(\Phi_2 + \Delta)}{\tilde{F}(\Phi_2)}
\right]
\tilde{H}(\Phi_2 + \Delta, \Phi_2),
\quad
\label{dotDelta}
\end{eqnarray}
where $\tilde{F}(\Phi_2) = F(\phi_2)$
and $\tilde{H}(\Phi_1, \Phi_2) = H(\phi_1, \phi_2)$.
We now average Eq. (\ref{dotDelta}) over the period $0 < t < T$,
assuming that $\Delta$ on the RHS is constant over a cycle,
which is justified to the leading order in the coupling $H$~\cite{Kuramoto}.
We thus obtain the evolution equation in the form of $\dot\Delta = -V'(\Delta)$ with
an effective potential $V(\Delta)$. The potential is calculated and plotted
in Fig. \ref{fig:VofDelta} for two examples, which are both non-rotated (${\bf Q}_i = {\bf I}$)
and in the far-field ($d \gg b$).
For the circular trajectory, the coupling function $H$
contains only the second harmonic as a function of $\phi$,
which results in either the in-phase ($\Delta=0$)
or anti-phase ($\Delta=\pi$) synchronization
depending on the force profile, as illustrated in
Fig. \ref{fig:VofDelta}(a).
For a more complex trajectory like the ellipse,
more than one stable and/or metastable solutions
can be obtained, as shown in Fig. \ref{fig:VofDelta}(b).
Note that the non-zero values of $\Delta$ generally means
non-constant phase shift $\delta = \phi_1 - \phi_2$
in the original gauge, as Fig. \ref{fig:VofDelta}(c) shows.

\commentoff{
Remark: we can simplify Eq. (\ref{dotDelta}) if the variation of $F(\phi)$
is small. Putting $F(\phi) = F_0 [1+f(\phi)]$ where $F_0$ is the average force
and  $f(\phi)\ll 1$, we get $\Phi(\phi) \simeq \phi$ and
\begin{eqnarray}
\dot{\Delta} \simeq -\frac{1}{T} \int_0^{2\pi} d\phi [f(\phi + \Delta) - f(\phi)] H(\phi+\Delta, \phi)
= -V'(\Delta)
\end{eqnarray}
For example, for the circular trajectory we have $H(\phi+\Delta, \phi) \propto  -\cos(2\phi+\Delta)$
and it couples with only the second harmonic of $f(\phi)$.
For $f(\phi) \propto \sin(2\phi + \delta_0)$, where $\delta_0$ is a phase delay,
we have $f(\phi + \Delta) - f(\phi) = 2 \cos(2\phi + \Delta + \delta_0) \sin\Delta$
and thus $V'(\Delta)$ is identically vanishing only if $\delta_0 = \pm \pi/2$.
When $\delta_0 \neq \pm \pi/2$, the function $V'(\Delta)$ is proportional to $\sin\Delta$
and is vanishing only if $\Delta = 0,\pm\pi$.
More generally in the far-field limit,
we have $H(\phi+\Delta,\phi) \propto -\cos(2n\phi+n\Delta)$
if $t_x(\phi) \propto \sin(n\phi)$, and
further if $f(\phi)$ contains the $n$-th harmonic,
then $V'(\delta) \propto \sin (n\Delta/2)$
and hence $\Delta=2\pi k/n$ ($k=0,1,2,n-1$) gives the stable solutions.
However, it is not quite likely that $t_x(\phi)$ contains only the $n$-th harmonic with $n\ge 3$,
because ${\bf t}(\phi)$ is a $2\pi$-periodic function by definition.
Let us consider the case where $t_x(\phi) = -\sin\phi (1 + B/2 \cos2\phi)$, $B\ll 1$.
Then $H(\phi+\Delta,\phi) \propto -\cos(2\phi+\Delta)
- B[\sin(\phi+\Delta) \cos(2\phi) + \sin(\phi) \cos(2\phi+2\Delta)]$
}

\commentoff{
For example, for the circular trajectory with the force profile
$F(\phi)=  F_0 [1 - A \sin (2\phi)]$,
we have $\Phi(\phi) = \frac{2\pi \zeta b}{T F_0} K(\phi)
= 2\pi K(\phi)/K(2\pi)$,
where $K(\phi) = \int_0^\phi d\phi'/[1 - A \sin (2\phi')]$.
The cycle average is taken by using $\Phi_2 \simeq 2\pi t/T$
and $H(\phi_1, \phi_2) = G_A \cos(\phi_1-\phi_2) + G_B \sin\phi_1 \sin\phi_2$
in Eq. (\ref{dotDelta}).
}

\paragraph{Synchronization near a substrate.}
Finally, let us consider the case where the rotors are suspended
at height $h$ from a flat substrate (located at $z=-h$).
In this case, the hydrodynamic coupling is expressed by
the Blake tensor~\cite{Blake}, which takes into account the no-slip boundary
condition on the substrate.
For simplicity, we restrict ourselves to the case where the trajectories
are confined in the $xy$-plane and are separated by a relatively large
distance, namely $d \gg h,b$. In this case, we can use Eq. (\ref{G12})
with $G_I(r) = 0$, $G_D(r) = 9h^2/r^3$. While all the results discussed
above hold true within the above restriction, it is straightforward
to extend the analysis to more general and complex geometries.

\paragraph{Discussion.}

Our analysis shows that the requirement for hydrodynamic synchronization
is nontrivial but not difficult to meet, and that a wide variety of beating
patterns do induce synchronization. For the example of circular trajectories,
we found that the logarithm of the force should contain
second-harmonic component of a specific sign.
This can be met if the force profile has
the second harmonic directly, or, via frequency doubling, if it has
the first harmonic. However, force patterns that only have harmonics
higher than two cannot synchronize. 
Dependences on the trajectory shape and geometry can be summarized
by representing the action of a rotor at far distance
by force multipoles at a fixed position.
For circular orbits, $\Gamma$ is independent of the size of the 
trajectory, which indicates that the dominant interaction comes from
force monopoles when they are in bulk fluid, or dipoles near a substrate
(each made of the force monopole and its mirror image~\cite{Blake}).
Linear oscillators do not couple at the first order 
of the force-multipole expansion. 
Their leading order coupling comes from the monopole-dipole interaction 
for beads that are in bulk, or dipole-quadrupole interaction near a substrate.
Linear oscillators are also special in the sense that they do not synchronize
if they are parallel. This could be relevant to the synchronization
of microswimmers~\cite{PY09}, which can be minimally modeled
by a linear configuration of three point forces~\cite{RG+AA-08}.
We have also derived a fully nonlinear
evolution equation for the phase difference, which enables us to study 
the global stability of synchronized states with or without phase shifts.
We note that the presence of stable phase shifts has been recently observed
in experiments on the beating patterns of the flagella of
{\em C. reinhardtii}~\cite{Goldstein}.
It is not difficult to extend the present analysis to 
helices (as a model of flagella) or other extended objects
to make comparison with experiments.

In conclusion, we have derived a generic and explicit criterion
for the trajectory shape and force profile that stabilize
synchronized states.
The criterion could be helpful in understanding
the collective behavior of active biological organelles,
and designing active microfluidic components
that could be tuned in and out of synchronized states
using mechanical signals
communicated via hydrodynamic interactions.

\acknowledgments
RG would like to acknowledge financial support from the EPSRC.


\begin{thebibliography}{99}
\bibitem{note_on_zeta}
The drag coefficient is modulated in the presence of the
substrate. However, the correction is of $O(a/h)$ and
is negligible in our leading order calculations.
\end{thebibliography}
\end{document}